\documentclass[12pt,preprint]{aastex}
\usepackage{graphicx,color}

\shorttitle{Heating Mechanisms in a Supra-arcade Plasma Sheet} 
\shortauthors{Reeves et al.} 

\begin{document}

\title{An exploration of heating mechanisms in a supra-arcade plasma sheet formed after a coronal mass ejection }

\author{Katharine K. Reeves\altaffilmark{1}, Michael S. Freed\altaffilmark{2}$^,$\altaffilmark{3}, David E. McKenzie\altaffilmark{2}$^,$\altaffilmark{4} and Sabrina L. Savage\altaffilmark{4}}
\email{{\tt kreeves@cfa.harvard.edu}}

\affil{$^1$Harvard-Smithsonian Center for Astrophysics, 60 Garden St. MS 58, Cambridge, MA 02138, USA}
\affil{$^2$Montana State University, Bozeman, MT 59717, USA}
\affil{$^3$Radford University, Radford, VA 24142, USA}
\affil{$^4$NASA Marshall Space Flight Center, Huntsville, AL 35812, USA}

\begin{abstract}
We perform a detailed analysis of the thermal structure of the region above the post-eruption  arcade for a flare that occurred on 2011 October 22.  During this event, a sheet of hot plasma is visible above the flare loops in the 131 \AA\ bandpass of the Atmospheric Imaging Assembly (AIA) on the {\it Solar Dynamics  Observatory (SDO)}.  Supra-arcade downflows (SADs) are observed traveling sunward  through the post-eruption plasma sheet.  We calculate differential emission measures using the AIA data and derive an emission measure weighted average temperature in the supra-arcade region.  In areas where many SADs occur, the temperature of the supra-arcade plasma tends to increase, while in areas where no SADs are observed, the temperature tends to decrease.  We calculate the plane-of-sky velocities in the supra-arcade plasma and use them to calculate  the potential heating due to adiabatic compression and viscous heating.  Ten of the 13 SADs studied  have noticeable signatures in both the adiabatic and the viscous terms.  The adiabatic heating due to compression of plasma in front of the SADs is on the order of 0.1 - 0.2 MK/s, which  is similar in magnitude to the estimated conductive cooling rate.  This result supports the notion that SADs contribute locally to the heating of plasma in the supra-arcade region.  We also find that in the  region without SADs, the plasma cools at a rate slower than the estimated conductive cooling, indicating additional heating mechanisms may act globally to keep the plasma temperature high.
 \end{abstract}

\section{Introduction}

Hot sheets of plasma above post-eruption arcades (historically referred to as ``fans of coronal rays'') were 
first noticed using data from the Soft X-ray Telescope (SXT) on {\it Yohkoh} \citep{Svestka1998}.  Early 
observations of this region showed downflowing voids in the plasma, currently referred to as supra-arcade 
downflows \citep[SADs;][]{McKenzie1999,Innes_b2003,Innes_a2003}.  With the introduction of the Atmospheric 
Imaging Assembly (AIA) on the {\it Solar Dynamics Observatory (SDO)}, many of these plasma sheets have been 
observed with the 131 \AA\ channel \citep[e.g.][]{Reeves2011,Warren2011,Savage2012,LiuW2013, Hanneman2014, 
Innes2014,Gou2015,Guidoni2015,LiL2016}, which is sensitive to plasma at about 10 MK \citep{Boerner2012}.

Much of the work on the thermal structure of supra-arcade plasma in the low corona has focused on cusp-shaped 
structures above flare loops.  Using {\it Yohkoh}/SXT data, \citet{Tsuneta1996} found that a cusp-shaped flare 
had hot plasma on the outer edges of the cusp, with a channel of cooler, denser plasma down the middle.  
\citet{Gou2015} surveyed many cusp-shaped flares using AIA and found that most of them adhered to this 
pattern.  \citet{Guidoni2015} study a flare using AIA data and find a similar temperature structure, but 
without the cooler channel.  In the cusp-shaped event studied by \citet{LiuW2013}, the high temperature region 
is always below the reconnection site, indicating that the heating is due to the reconnection outflow.

Very little work has been done on the temperature structure of the supra-arcade downflows themselves, but 
\citet{Hanneman2014} examined the SADs in several flares, and found that for the most part they are cooler than 
the surrounding fan plasma.  \citet{LiL2016} also found that a SAD embedded in a plasma sheet is cooler than 
its surroundings.

The mechanism that heats the supra-arcade plasma is not fully understood.  \citet{Tsuneta1996} speculated that 
the cusp could be heated by standing isothermal slow shocks emanating from the reconnection region.  Models 
show that heating from compression can also play a role \citep{Birn2009}, and thermal conduction can spread 
heating from ohmic dissipation in the current sheet to form a ``thermal halo'' of hot plasma 
\citep{YokoyamaShibata2001, Seaton2009, Reeves2010}.  There is some evidence for conduction-broadened regions 
around current sheets in observations.  For example, \citet{Hannah2013} observed an eruption at the limb that 
consisted of a hot (8-14 MK) plasmoid connected to the flare arcade by a ``stem'' of hot material which is wider 
than a theoretical current sheet.  \citet{Savage2010} measured the width of a post-CME plasma sheet observed by the X-ray Telescope (XRT) on {\it Hinode} to be on the order of 5$\times$10$^3$ km, and current sheet like structures have been found with thicknesses 10-100 times greater this value at heights of $>$1.3 $R_{sun}$ \citep{Ciaravella2002,Ko2003,Webb2003,Lin2007}. 

In this article, we investigate mechanisms due to the motion of the SADs that could have a local heating effect 
on the supra-arcade plasma, i.e. adiabatic compression and viscous heating.  The supra-arcade downflows are 
tracked using a semi-autonomous tracking algorithm \citep{McKenzie2009}, and plasma velocities are calculated 
using a local correlation tracking algorithm \citep[as in][]{McKenzie2013, FreedMcKenzie2016}.  Temperatures are 
estimated from differential emission measures calculated from the AIA observations.  Section 2 presents the AIA 
observations used in this work.  Section 3 presents the temperature calculations for the supra-arcade plasma and 
the calculation of the heating terms in the same region. The discussion is presented in Section 4, and the 
conclusions are given in Section 5.
  
\section{Observations \label{observations.sec}}

\begin{figure}
\includegraphics[scale=0.4]{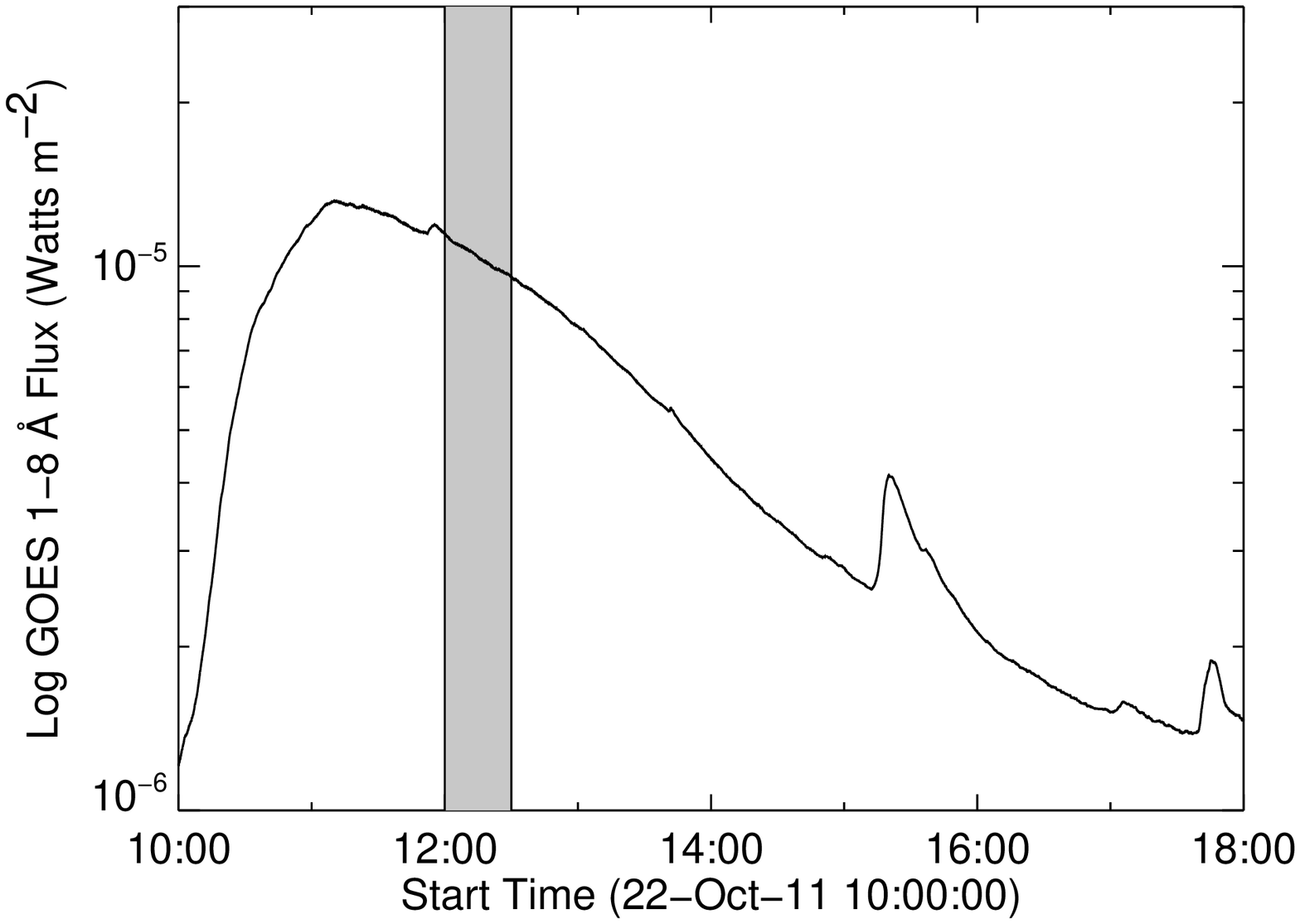}
\caption{\label{goes.fig}  GOES plot depicting the soft X-ray flux for the long duration even that occurred on 2011 October 22.  The shaded region indicates the time period during which we examine the thermal structure of the supra-arcade fan in relation to the presence of downflows.}
\end{figure}

On 2011 October 22, starting at about 10 UT, a flare of magnitude M1.3 erupted on the northwest limb of the Sun.  
The peak X-ray intensity was achieved at 11:10 UT.  The flare was a long duration event, with the decay phase 
lasting for several hours after the peak flux.  A GOES plot of the event is shown in Figure \ref{goes.fig}.  The 
flare was also observed by the Atmospheric Imaging Assembly \citep[AIA;][]{Lemen2011} on the Solar Dynamics 
Observatory (SDO). 

\begin{figure*}
\includegraphics[scale=0.8]{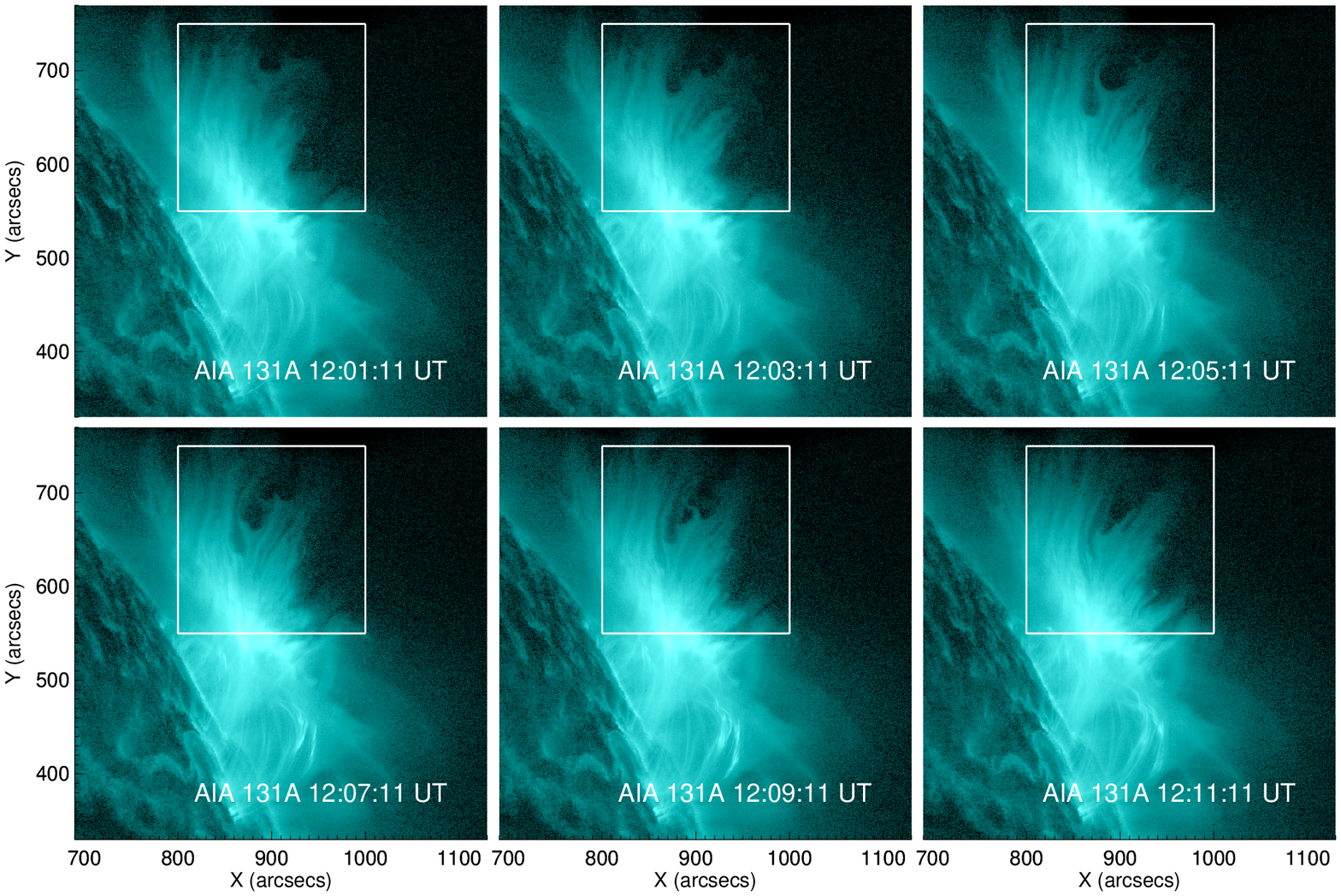}
\caption{\label{summary.fig} Evolution of the supra-arcade downflows in the plasma sheet formed after the long 
duration M1.3 flare that peaked at 11:10 UT on 22 October 2011.  The white box shows the area of interest for 
later figures.}
\end{figure*}

 We choose this particular flare for our analysis because of the numerous, large SADs that occur after the eruption.
During the flare's long decay, these SADs are seen clearly in the AIA 131 
\AA\ channel, as shown in Figure \ref{summary.fig}.  The morphology of the SADs in this event was studied 
previously by \citet{Savage2012}, who interpreted the areas of low intensity in the 131 \AA\ channel as voids 
that were preceded by shrinking loops.  \citet{Hanneman2014} also studied this event, and found that in three 
of the largest SADs in this flare, the temperature in the SAD was lower than the temperature in the surrounding 
fan plasma. The large size of the SADs in this event allows us to clearly follow changes in the plasma surrounding them.

For this investigation, we primarily use images from the AIA EUV channels, which are taken with a cadence of 12 
s.  AIA has a full-sun field of view, and a spatial resolution of $\sim$0.6\arcsec per pixel.  The AIA data 
are prepped using the {\tt aia\_prep} routine, available in the SolarSoft Ware 
\citep[SSW;][]{Freeland1998} package.  This routine de-rotates the AIA images from the four telescopes, aligns 
them, and re-factors the images so that they all have the same plate scale.  We deconvolve the images with a 
standard point-spread function using the {\tt aia\_deconvolve\_richardsonlucy} routine, also available in the 
SSW package.

\section{Results \label{results.sec}}

\subsection{Downflow tracking}

In order to find the paths traversed by the SADs as they descend through the fan, we use the semi-automated 
tracking method developed by \citet{McKenzie2009}.  In order to apply this algorithm, the data are further 
processed in order to enhance contrast and reduce background by differencing the images from a running mean.  The 
search and tracking algorithm identifies groups of pixels with locally decreased intensity in these processed images, and the 
centroid location of this region is recorded as the position of the SAD path at each time step.

\begin{figure}
\includegraphics[scale=0.5]{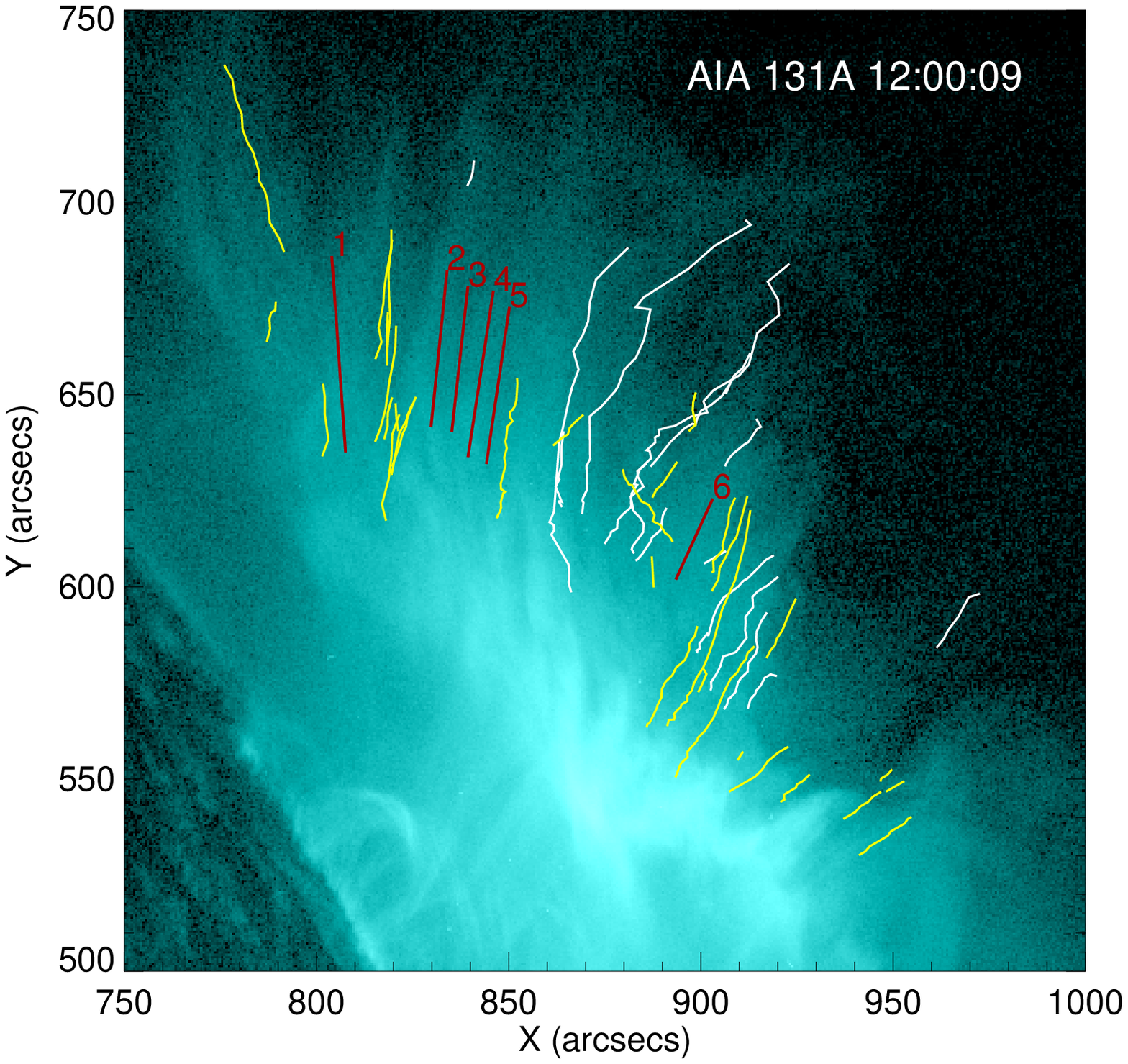}
\caption{\label{SAD_tracks.fig} Paths followed by the automatically (white) and manually (yellow) tracked 
downflows from 11:00 - 12:30 UT according to the algorithm developed by \citet{McKenzie2009}.  Lanes that are 
free of SADs during the same time period are marked with straight red lines.  (An animation of this figure is available in the online journal.)}
\end{figure}

We follow the SADs in the time period between 11:00 UT and 12:30 UT, and confirm that there are both SADs and 
supra-arcade downflowing loops (SADLs) during this time period, as was found by \citet{Savage2012}.  The paths 
taken by the SADs are shown overplotted on an AIA 131 \AA\ image in Figure \ref{SAD_tracks.fig}, and it is clear 
from this figure that the SADs (tracks shown in white in Figure~\ref{SAD_tracks.fig}) and SADLs (tracks 
shown in yellow in Figure~\ref{SAD_tracks.fig}) are more prevalent in some areas of the fan.  For example, 
the left side of the fan is dominated by SADLs, and there are no SADL tracks in an area centered around 
(840\arcsec,650\arcsec).  Examination of the movie provided with Figure \ref{SAD_tracks.fig} shows that while 
the fan plasma is somewhat turbulent in this region, there are no identifiable SADs or shrinking loops 
during the time frame in question.  

There is another lack of tracked downflows centered around 
(930\arcsec,570\arcsec), but examination of the movie provided with Figure \ref{SAD_tracks.fig} shows that there 
are some SADs not identified by the tracking algorithm that occupy this region starting at around 12:15 UT 
This oversight is due to the high detection threshold set when applying the procedure to a region of relatively low fan intensity.
There is also a small gap in tracked downflows centered at about 
(900\arcsec,610\arcsec), where there is a solitary red line in Figure \ref{SAD_tracks.fig}.  The straight red lines in Figure 3 indicate lanes with no clear indications of downflows, except for the aforementioned suggestion of turbulent motion.
 
\subsection{Thermal properties of supra-arcade plasma}

We use the routine {\tt xrt\_dem\_iterative2} \citep{Weber2004, Golub2004} to calculate the DEMs in the region 
indicated by the white boxes in Figure \ref{summary.fig}.  This routine finds the DEM by manipulating spline 
knots in DEM space until the $\chi^2$ between the simulated intensities and the observed intensities is 
minimized.  This routine has been extensively tested using model DEMs \citep{Cheng2012}, and has been shown to 
provide similar results as other algorithms \citep{ SchmelzSaar2009,SchmelzKashyap2009,HannahKontar2012}.

We calculate the DEMs in every pixel, except for those deemed to have too weak a signal in the AIA 131 \AA\ 
channel to use the flow tracking algorithm (described in the next section).  The DEMs are calculated from the 
intensities in the six AIA EUV channels that are predominantly dominated by emission from iron lines (94 \AA, 131 \AA, 171 \AA, 193 \AA, 211 \AA\ and 335 \AA).  We then 
calculate an average emission-measure weighted temperature, given by

\begin{equation}
\label{t_em.eq}
T_{em} = \frac{\int T\times DEM(T)dT}{\int DEM(T)dT}
\end{equation}
and a total emission measure, given by
\begin{equation}
EM_{tot} = \int DEM(T)dT.
\end{equation}

Typical DEMs for the supra-arcade fan in this flare consist of a low temperature component at logT $\simeq$ 6.2 
and a higher temperature component at logT $\simeq$ 7.0 \citep[see Figure 5 in][]{Hanneman2014}.  Some studies 
have suggested that integrating only the high temperature component in Equation \ref{t_em.eq} provides a better 
characterization of the temperature of the supra-arcade plasma \citep[e.g.][]{Gou2015}.  However, we elect to 
integrate over the entire DEM because we are also interested in the temperature of the voids associated with the 
SADs.

\begin{figure*}
\includegraphics[scale=0.4]{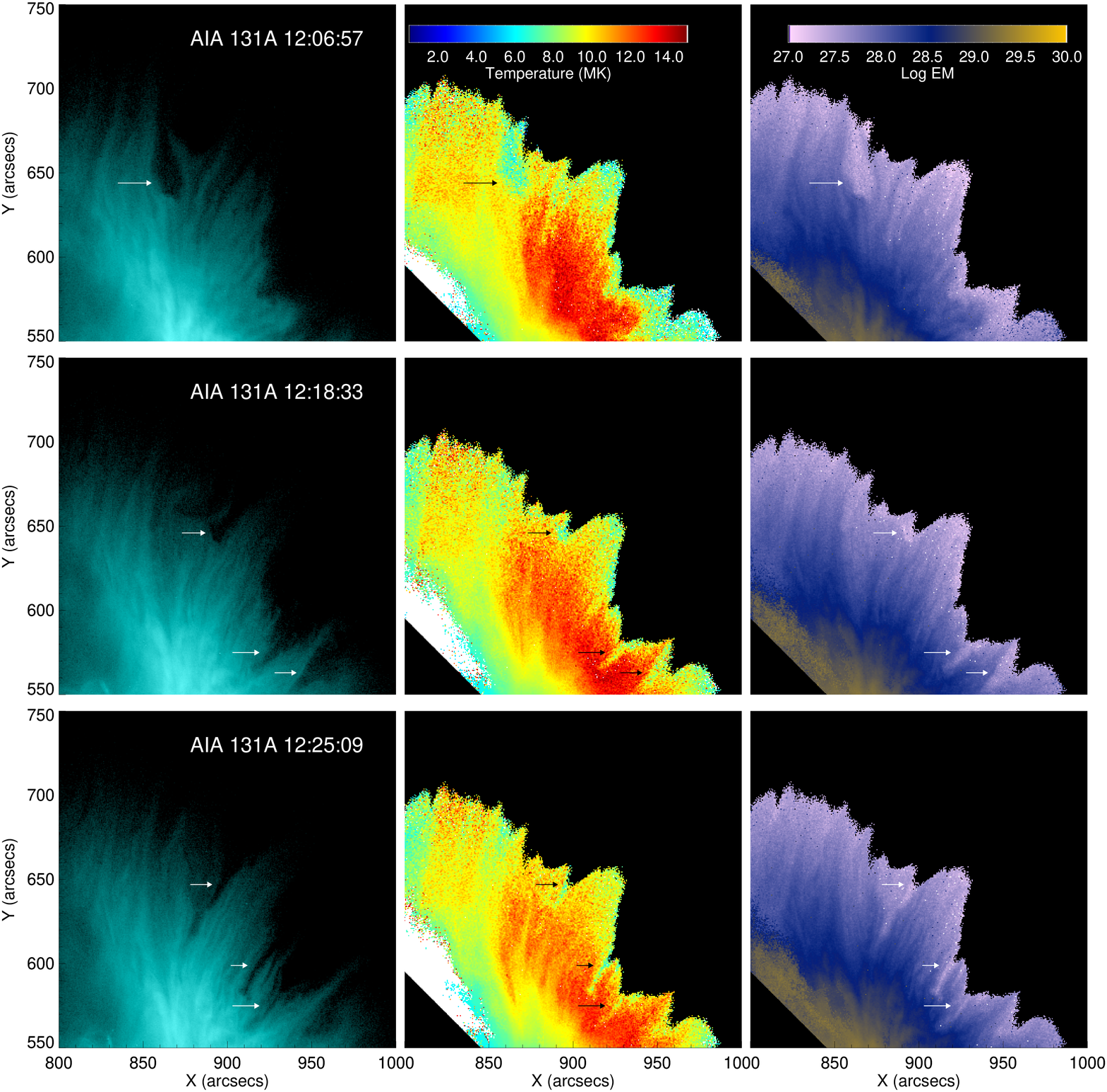}
\caption{\label{inten_temp_em.fig} Intensity in the AIA 131 \AA\ bandpass (left column), $T_{em}$ (middle column), and Log$EM_{tot}$ (right column) for several different times during the decay phase of the flare.  SADs are marked with arrows. (An animation of this figure is available in the online journal.)}
\end{figure*}

Figure \ref{inten_temp_em.fig} shows the AIA 131 \AA\ intensity, $T_{em}$, and $EM_{tot}$ for several times 
during the period of interest.  The large SAD seen in the images from 12:06:57 UT located at 
(870\arcsec,650\arcsec) (indicated by an arrow in the top row of Figure \ref{inten_temp_em.fig}) was one of the 
ones studied by \citet{Hanneman2014}.  Consistent with their results, the map of $T_{em}$ indicates a decreased 
average temperature in this region compared to the surrounding fan plasma.  The $EM_{tot}$ in this region is 
also depressed compared to the surrounding plasma.  Other SADs appear later in the evolution of the supra-arcade 
plasma, indicated by arrows in Figure \ref{inten_temp_em.fig}, and they generally follow the same trend, with 
the temperature and density inside the SAD depressed with respect to the surrounding fan plasma.   

The temperature and emission measure measurements reported here are consistent with previous interpretations of SADs as motions of cool, rarefied plasma descending through a hotter plasma sheet toward the flare arcade. It is conceivable that temperature variations within the plasma could mimic motion, via the plasma transitioning into and out of a given instrumentÕs temperature-dependent response passband, but this explanation is not consistent with other observational properties of SADs.  We note that SADs are frequently observed by instruments with much broader temperature response curves than AIA (e.g., {\it Yohkoh}/SXT, {\it Hinode}/XRT, RHESSI), and have displayed Doppler shifts difficult to explain away as temperature/passband effects \citep{Innes_b2003}.  \citet{FreedMcKenzie2016} have recently demonstrated conclusively that SADs observed simultaneously with {\it SDO}/AIA and {\it Hinode}/XRT can be interpreted as a temperature/passband effect only by invoking extreme, even ``pathological'', heating and cooling mechanisms operating on very short time scales.

\begin{figure*}
\includegraphics[scale=0.4]{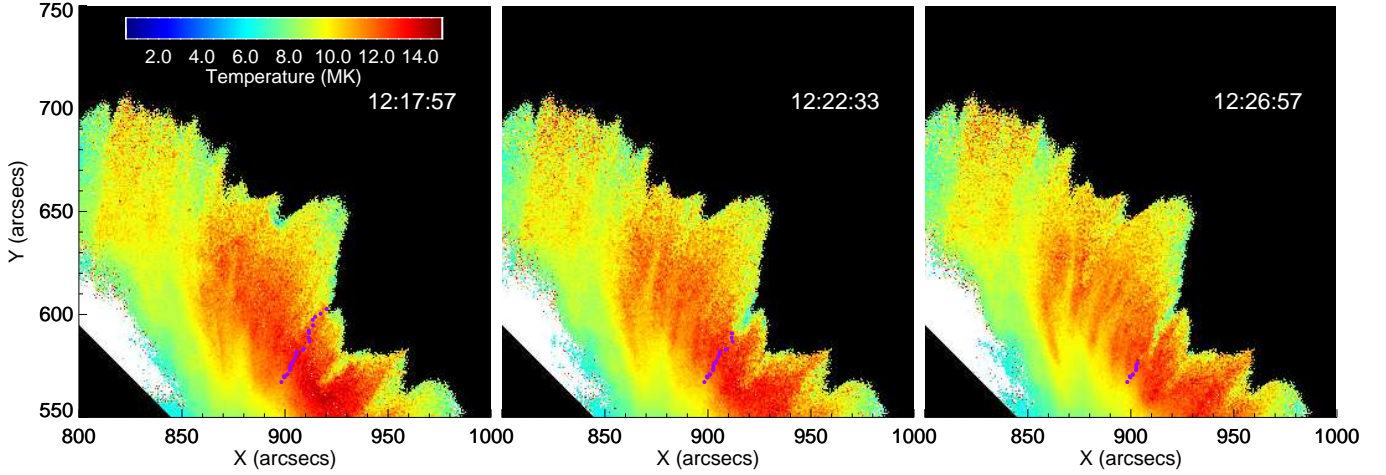}
\caption{\label{temp_tracks.fig} $T_{em}$ maps for several different times.  Purple dots indicate locations used to find average temperature in front of SAD 11 at each time.  The dots are derived from the SAD track, and lie along the path the SAD will eventually follow. }
\end{figure*}

In order to understand the effect of the motion of the SADs on the temperature evolution of the supra-arcade plasma as a whole, we 
examine the evolution of the temperature in areas with and without SADs.  Since the SADs themselves appear as voids that are cooler than their surroundings, we look at the temperature of the plasma ahead 
(sunward) of the SADs.  We average the temperature along the track that the SAD follows, excluding the plasma 
inside the SAD itself.  Figure \ref{temp_tracks.fig} shows an example for one of the tracks (SAD 11).  The 
temperature at each time step is averaged over the purple dots in Figure~\ref{temp_tracks.fig}.  We also 
average the temperature over lanes in the supra-arcade fan where there are no evident 
SADs.  These lanes are shown as red lines in Figure \ref{SAD_tracks.fig}.

\begin{figure}
\includegraphics[scale=0.45]{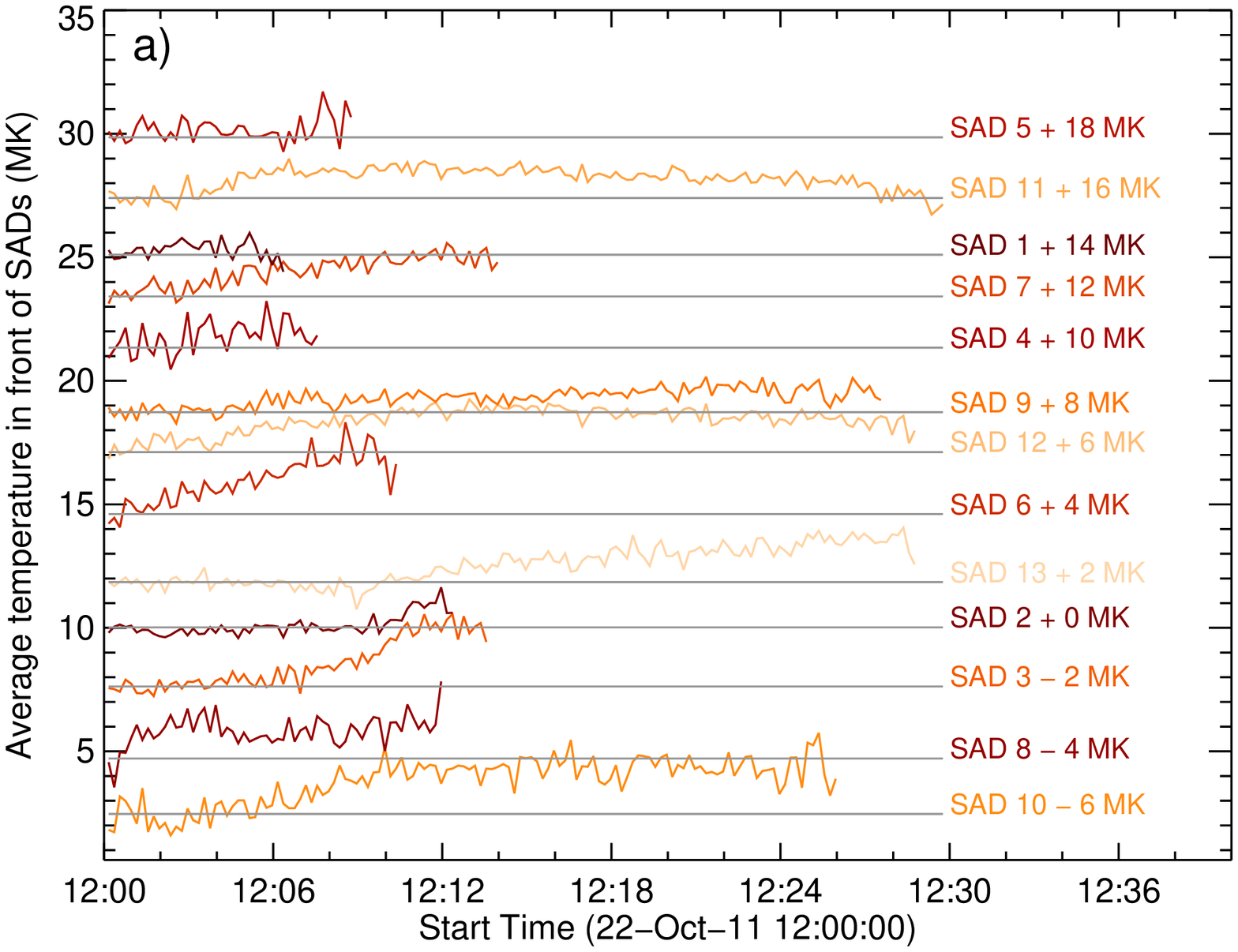}
\includegraphics[scale=0.45]{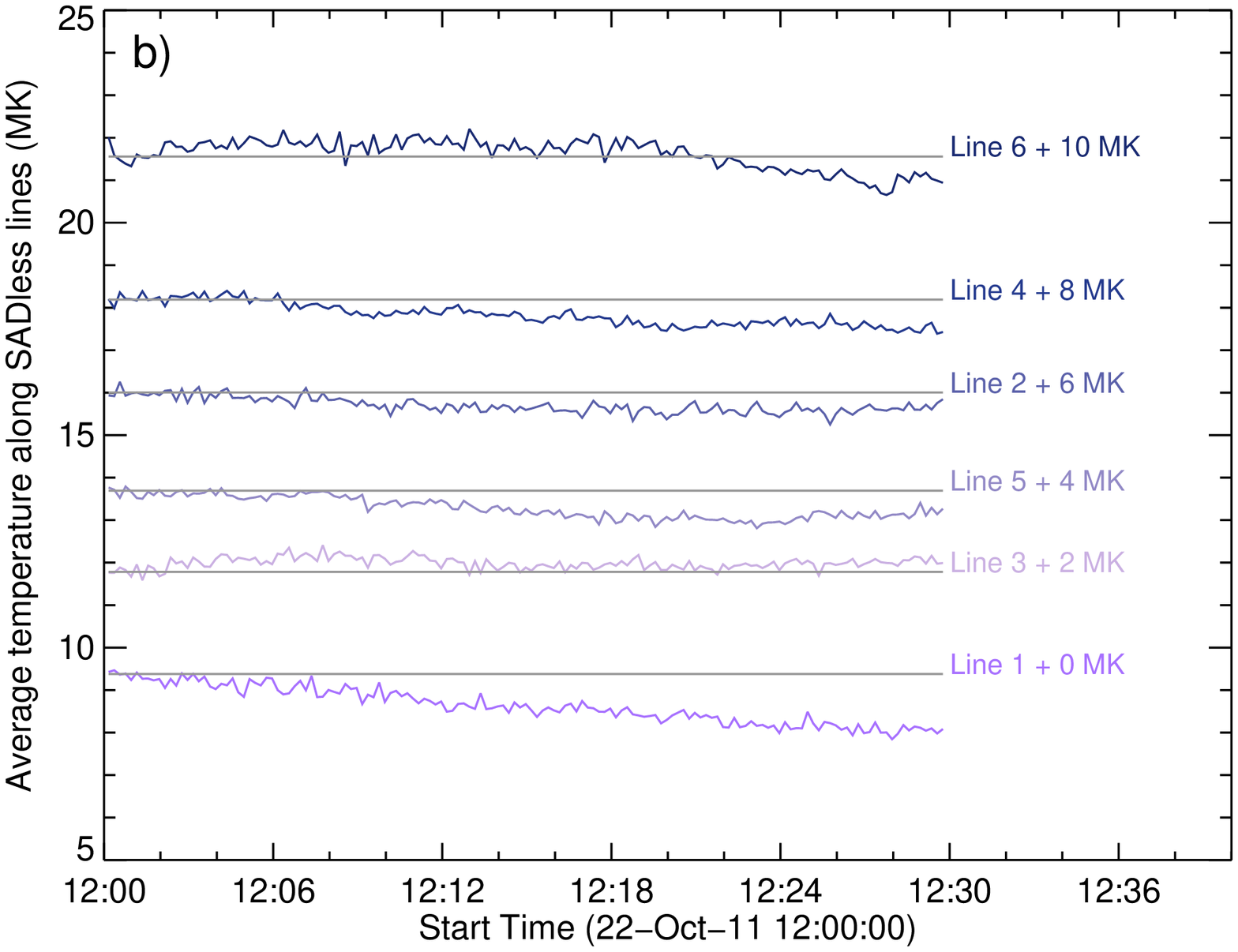}
\caption{\label{temp_trend.fig} Panel a) Average temperatures in front of the SADs (red lines) and along the 
lines with no SADs as a function of time.  Panel b) Same average temperatures as in panel a), but normalized to 
the initial average temperature. Temperatures have been shifted by the indicated amount for clarity.}
\end{figure}

Figure \ref{temp_trend.fig}a shows the average temperature in front of the SADs and Figure 
\ref{temp_trend.fig}b shows the average temperature in the regions with no SADs.  Also plotted are lines 
indicating the initial temperature for each feature, which is calculated by time-averaging the first five values 
of the average temperature.  Many of the SADs show an increase in temperature in front of the downflow, and some 
of the increases are on the order of 2MK or more, such as in the case of SAD 3 and SAD 6.  A few of the 
temperatures stay roughly the same in front of the SAD, as in SADs 1, 4, and 5.  None of the temperatures in 
front of SADs decrease markedly from their starting temperature.  On the other hand, for the lines in the area 
where there are no evident SADs, four of the six lines show a decrease in average temperature as a function of 
time (lines 1, 2, 4, and 5).  Line 6 maintains a relatively steady temperature until about 12:22 UT, when it 
begins to decrease.  Line 3 maintains a roughly constant temperature throughout the time period.
 
\subsection{Heating terms in the supra-arcade region}

As shown in Figure \ref{temp_trend.fig}, the plasma in front of the SADs tends to either stay the same temperature or heat up.  It does not cool down, as the plasma in the region without the SADs tends to.  In order to investigate the reason for this heating, we examine several mechanisms that may be responsible for heating the plasma.  
The MHD energy equation can be written in terms of temperature \citep[e.g.][]{Lionello2009}:

\begin{equation}
\label{energy.eq}
\left(\frac{\partial T}{\partial t} - {\bf v}\cdot\nabla T\right) = (\gamma-1)\left[-T\nabla\cdot{\bf v} +\frac{m}{k\rho}\mathcal{L}\right]
\end{equation}
where $\mathcal{L}$ contains sources and sinks of energy, including ohmic dissipation, conduction, radiation, viscous heating, and coronal heating.  Additionally, $\gamma = 5/3$ is the polytropic index, ${\bf v}$ is velocity, $k$ is Boltzmann's constant, $\rho$ is the mass density, and $m$ is the average mass per particle (for the Sun, we can take $m = 0.6m_p$, where $m_p$ is the mass of the proton.  See \citet{PriestMHDBook}, p.82).

Two terms that contribute to temperature changes in Equation \ref{energy.eq} are related to the plasma velocity, 
namely the heating due to viscous motions (which is contained within $\mathcal{L}$) and the adiabatic term.  The adiabatic term is given by:
\begin{equation}
H_a = -(\gamma - 1)T\nabla\cdot {\bf v}.
\end{equation}
Compressions (i.e. a negative $\nabla\cdot {\bf v}$) cause an increase in temperature, and rarefactions (i.e. a positive $\nabla\cdot {\bf v}$) cause a decrease in temperature.  The rate of viscous heating of the plasma is given by
\begin{equation}
H_\nu= \frac{(\gamma -1)m}{k\rho}h_\nu
\end{equation}
where $h_\nu$ is
\begin{equation}
\label{viscous.eq}
h_v = \rho\nu\left(2\left[\left(\frac{\partial v_x}{\partial x}\right)^2+\left(\frac{\partial v_y}{\partial y}\right)^2\right] + \tau(v_x,v_y)^2 - \frac{2}{3}(\nabla\cdot\mathbf{v})^2\right).
\end{equation}
In the above equation,  $\nu$ is the kinematic viscosity and $\tau(v_x,v_y)$ is the velocity shear, given by
\begin{equation}
\tau(v_x,v_y)= \left[\left(\frac{\partial v_x}{\partial y}\right) + \left(\frac{\partial v_y}{\partial x}\right)\right].
\end{equation}
For the kinematic viscosity, we use
\begin{equation}
\nu = 1.02\times 10^{8} \frac{T^{5/2}}{n} {\mbox cm^2 s^{-1}}
\end{equation}
where $n$ is the number density.  We calculate $n$ from the emission measures in each pixel derived in the previous section using the formula 
\begin{equation}
EM = n^2l
\end{equation}
where we assume a path length $l=10^9$ cm.

Velocities in the supra-arcade plasma sheets are calculated by applying Fisher and Welsch's Fourier Local Correlation Tracking program \citep[FLCT:][]{Fisher2008} to contrast-enhanced AIA images. The uncertainties associated with the FLCT velocities are described in \citet{Freed2016}, but in general, results are found to consistently underestimate the actual velocities \citep{Svanda2007,Verma2013,Loptien2016}. The FLCT parameters' fidelity were determined by applying the cork-advection-method used by \citet{McKenzie2013}. This method superimposes a series of virtual particles, ÒcorksÓ, advected by the FLCT velocities with their corresponding AIA intensity images in the background. The FLCT parameters, described in \citet{Fisher2008}, are adjusted until the best agreement is found between the corks and intensity images. Using FLCT in this manner has been previously used to track flows in plasma sheets found in prominences \citep{Freed2016} and the supra-arcade region above flare loops \citep{FreedMcKenzie2016}.

As is evident from Equation \ref{viscous.eq}, we assume that the plasma motions are purely in the $x$-$y$ plane (i.e. the plane of the sky).  This assumption is justified because supra-arcade plasma sheets tend to be quite narrow when observed edge on \citep[e.g.][]{Savage2010,LiuR2013, LiuW2013, Gou2015}, limiting the ability of the plasma to flow in the direction perpendicular to the plasma sheet.  The neutral line of the active region that produces this eruption is very close to being aligned with the North-South direction, as can be seen in images from {\it STEREO-A}/SECCHI, which was positioned ahead of the Earth with a separation angle of about 105 degrees \citep[see Figure 2 in][]{Savage2012}.  Thus the measured plane-of-sky motions are a reasonable approximation for the velocities within the plasma sheet.   Though a plane-of-sky approximation is reasonable for this event, we note that data of the supra-arcade region from a fast-cadence spectrometer, such as the {\it Interface Region Imaging Spectrograph} \citep[{\it IRIS},][]{DePontieu2014}, would be very useful for determining the line-of-sight velocities, and thus determining the flow field more accurately.

 \begin{figure*}
\includegraphics[scale=0.35]{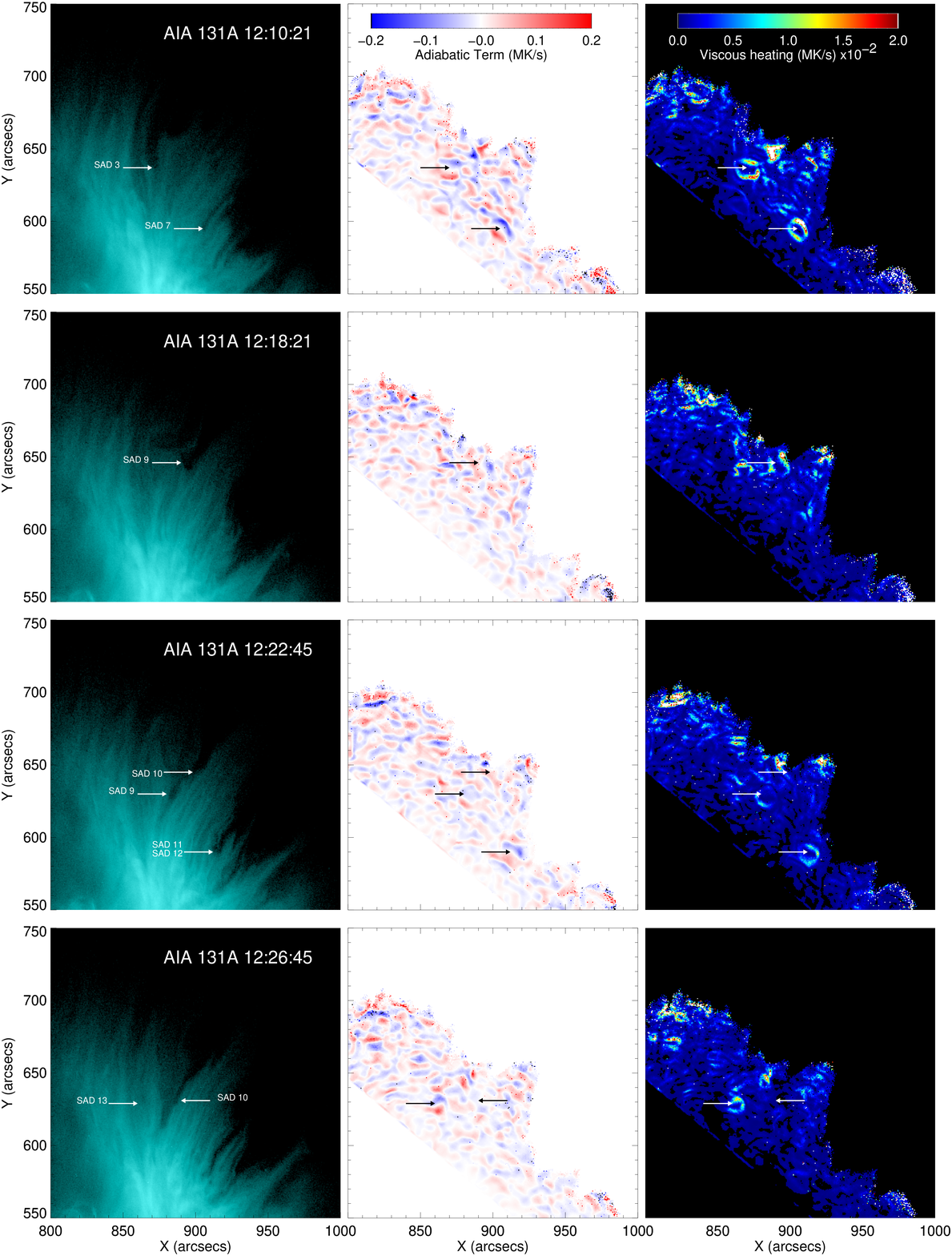}
\caption{\label{heating_terms.fig} Intensity in the AIA 131 \AA\ bandpass (left column), the adiabatic term $H_a$ (middle column), and the viscous heating term $H_{\nu}$ (right column) for several different times during the decay phase of the flare.  SADs are marked with arrows. (An animation of this figure is available in the online journal.)}
\end{figure*}
 
Figure \ref{heating_terms.fig} shows the AIA 131 \AA\ intensity, the adiabatic term and the viscous heating for 
several times during the post-eruption phase.   The viscous heating term shows clear signatures that correspond with the SADs, consisting of distinct ring shapes circling the heads of the SADs that travel sunward with the downflows.   For example, SADs 3 and 7, shown in the top panels of Figure \ref{heating_terms.fig} have ring signatures around the heads of the SADs that stand out quite clearly from the background. The second row of Figure \ref{heating_terms.fig} shows SAD 9,  which is traveling from right to left, rather than radially towards the sun so the ring structure in the viscous term is tilted.  SAD 9 is also shown in the third row of Figure \ref{heating_terms.fig} at a time when its motion has become more radial. At this time, the ring in the viscous term is still discernible, though it has faded in intensity.  Also included in this part of the figure are SADs 11 and 12, which descend in unison, right next to each other. They share a clear ring in the viscous heating term.   SAD 13, which appears in the fourth row of Figure \ref{heating_terms.fig}, also has a clear ring signature in the viscous term. This SAD is very narrow, and hard to see in the still images.

 The adiabatic term also shows discernible patterns around the heads of the SADs, though there is less contrast between these signatures and the background than there is with the viscous term.  Typically, there is an increase in   
the adiabatic term (indicated by a red color) in front of the SAD and a decrease (indicated by a blue color) 
behind the SAD, consistent with compressional heating in front of the SAD and cooling due to rarefaction behind 
it.  These patterns  are in the same locations relative to the SADs as the viscous signatures discussed in the previous paragraph.  In the top and bottom rows of Figure \ref{heating_terms.fig}, strong versions of these patterns associated with SADs 3, 7, and 13 can be seen.  A tilted version of this pattern can be seen in the second row of Figure \ref{heating_terms.fig}, accompanying SAD 9 as it moves from right to left.  This pattern is still visible around SAD 9 in the third row of Figure \ref{heating_terms.fig}, though the intensity has faded somewhat.  The third row of Figure \ref{heating_terms.fig} also shows a relatively strong alternating red/blue pattern around SADs 11 and 12.

 The patterns in the adiabatic heating term associated with the SADs are only at most a few times the background signal, but they are distinct from the background in that they travel with the SADs as they descend (see the animation accompanying Figure 
\ref{heating_terms.fig}).  The background signal in the adiabatic term may be due to noise amplification inherent in taking derivatives of the flow field, or it may be due to motions on smaller scales than the SADs.  Very fast and narrow SADs, for example, might create patterns that were less obvious than the ones we have identified, especially since these fast and narrow SADs would be difficult to identify with our tracking algorithm.

Not every SAD has associated patterns in the viscous and adiabatic terms.  The third row of Figure \ref{heating_terms.fig} shows SAD 10, which has no 
discernible red/blue pattern in the adiabatic term or ring in the viscous heating term.  The fourth row of 
Figure \ref{heating_terms.fig} shows SAD 10 at a later time, and again no red/blue pattern or ring is visible.  

A summary of the SADs and their correlated features is given in Table \ref{heating.tab}.  For most of the SADs 
tracked in this flare, there is a noticeable signature in the adiabatic term consisting of heating in 
front of the SAD and cooling behind it, and a noticeable ring of enhanced viscous heating surrounding 
the head of the SAD.  The exceptions are SADs 4, 8 and 10. SAD 4 does not travel very far, and also is not 
associated with a discernible temperature increase.  SAD 8 has a temperature increase right at the beginning of 
the time period studied, but remains steady afterwards.  SAD 10 descends in nearly the same place as SAD 9 did a 
few minutes earlier, so it is traveling through the rarefied plasma in the wake of SAD 9.  SADs 1 and 5 are 
interesting in that they both have identifiable signatures in the adiabatic and viscous terms, but don't exhibit 
any real temperature increase in the plasma in front of them.  These two SADs seem to collide, though, which may 
complicate the relationship between their motion and the plasma temperature.

\begin{table}
\begin{center}
\caption{\label{heating.tab}Summary of physical parameters for SADs}
\begin{tabular}{lllll}
\tableline\tableline
SAD & Temp & Adiabatic & Viscous & Comments  \\
No. & increase? & signature? & signature? & \\
\tableline
& & & & \\
1 & No & Yes & Yes & Collides with SAD 5 \\
2 & Yes & Yes & Yes & Very wide SAD \\
3 & Yes & Yes & Yes & Very wide SAD \\
4 & No & No & No & Short SAD track \\
5 & No & Yes & Yes & Collides with SAD 1 \\
6 & Yes & Yes & Yes & Fairly short track \\
7 & Yes & Yes & Yes & Strong adiabatic signature\\
8 & Yes & No & No &  Temp rise early\\
9 & Yes & Yes & Yes & Some sideways motion \\
10 & Yes & No & No & Same region as SAD 9 \\
11 & Yes & Yes & Yes & Same as SAD 12 \\
12 & Yes & Yes & Yes & Same as SAD 11 \\
13 & Yes & Yes & Yes & Very thin SAD \\
\tableline
\end{tabular}
\end{center}
\end{table}

\begin{figure}
\includegraphics[scale=0.45]{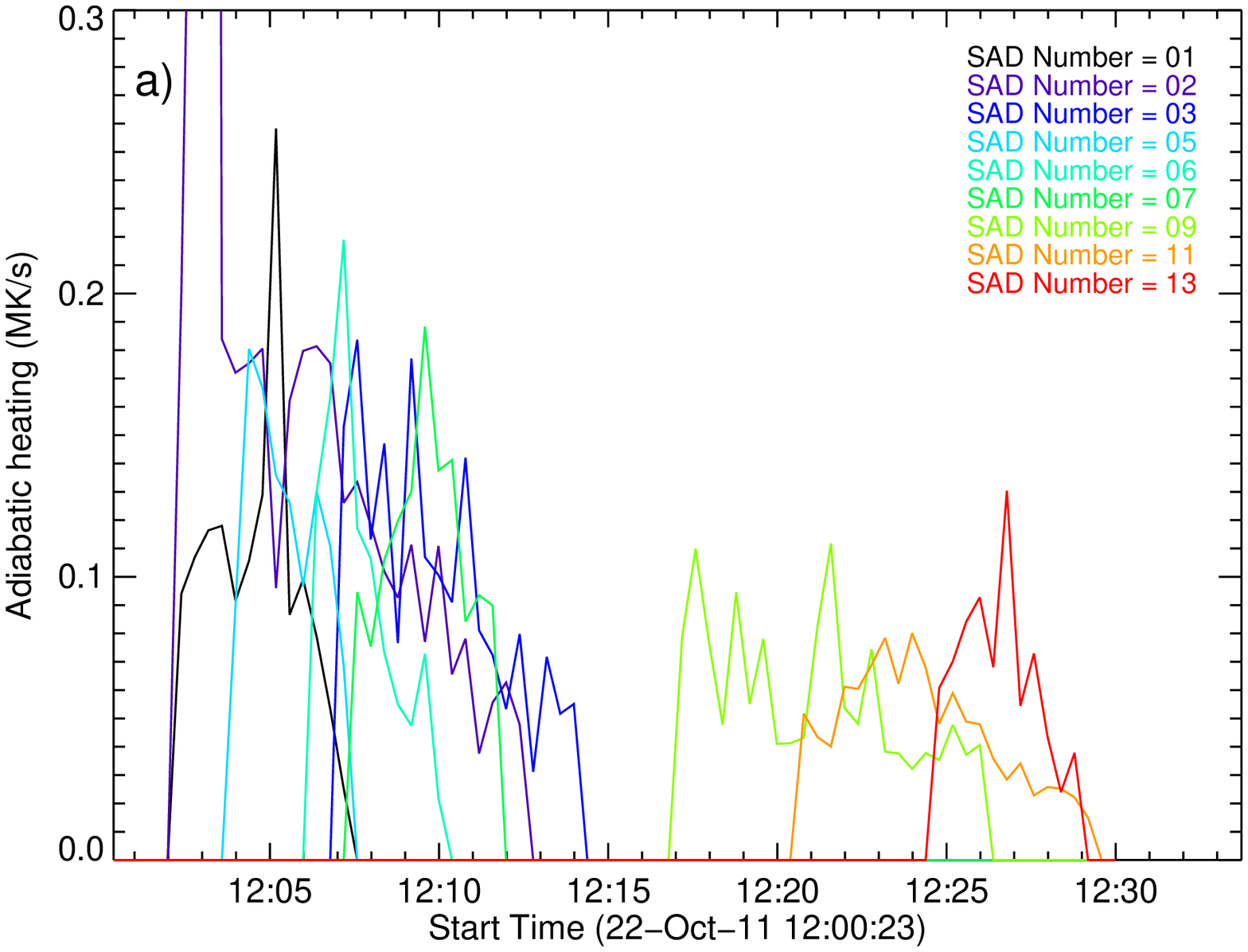}
\includegraphics[scale=0.45]{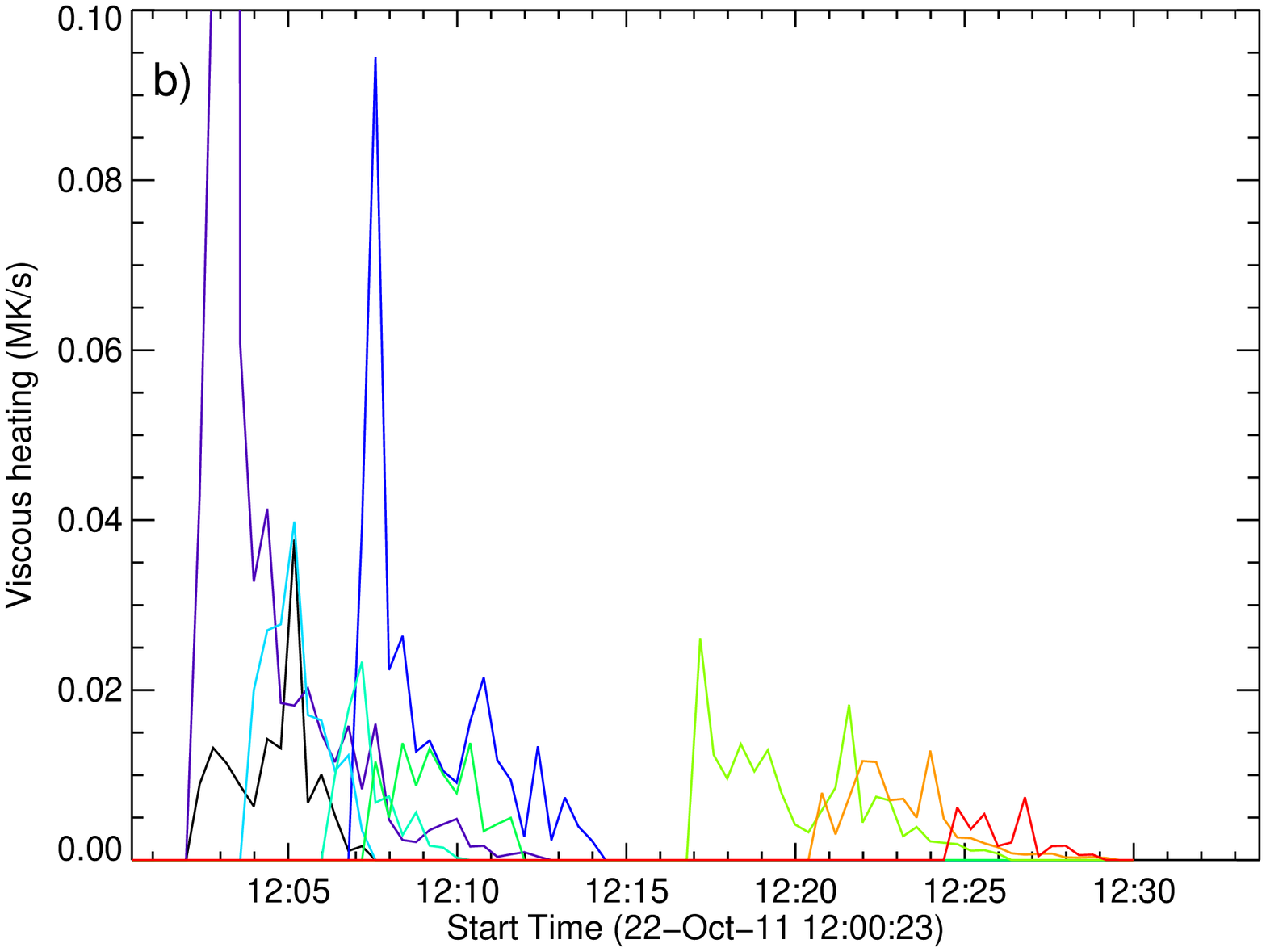}
\caption{\label{heating_values.fig}  Evolution of the values of adiabatic heating (panel a) and viscous heating (panel b) in front of the SADs that have heating signatures.}
\end{figure}

Figure \ref{heating_values.fig} shows the values of the adiabatic and viscous heating terms in front of the SADs where these signatures are visible.  The locations were selected manually, aiming for the highest value sunward of the descending SAD.  SAD 2 is very large spatially, and has the largest initial values of both heating terms, though the increase in temperature in front of SAD 2 shown in Figure \ref{temp_trend.fig} is not especially large.  SAD 3, on the other hand, has a larger temperature increase on the order of 2 MK, and it also has large values initially for both the adiabatic and viscous heating.  The SADs that travel through the supra-arcade region earlier tend to have larger adiabatic heating terms.  This trend is also somewhat visible in the viscous heating terms as well.

\section{Discussion \label{discussion.sec}}

\begin{figure*}
\includegraphics[scale=0.4]{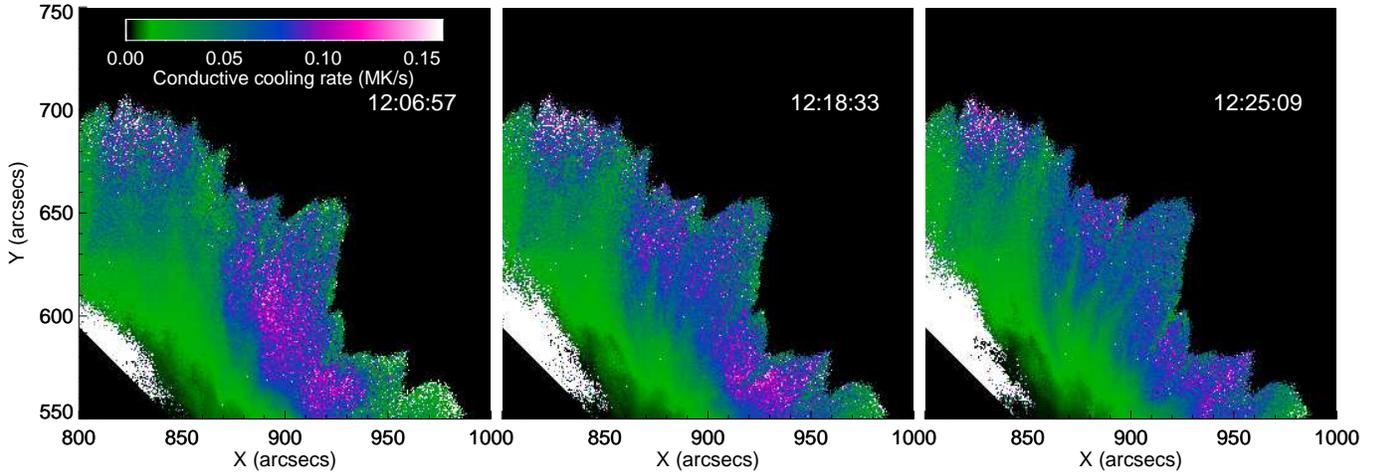}
\caption{\label{cond_cooling.fig} Estimated conductive cooling rates for the same times as shown in Figure \ref{inten_temp_em.fig}.}
\end{figure*}

As shown in Figures \ref{heating_terms.fig} and \ref{heating_values.fig}, the adiabatic heating rates associated with the SADs are on the order of 0.1-0.2 MK/s, and the heating rates due to viscous heating are an order of magnitude smaller than that.  The adiabatic term is clearly dominant over the viscous term, though the viscous term is easier to distinguish from the noise in the images in Figure \ref{heating_terms.fig}.  We want to determine if the adiabatic heating is sufficient to overcome the cooling of the supra-arcade plasma.  The supra-arcade fan plasma is at a temperature of about 10-15 MK, where conductive cooling should be very efficient.  The conductive cooling rate is given by
\begin{equation}
H_q = \frac{(\gamma -1)m}{k\rho}(-\nabla\cdot {\bf q})
\end{equation}
where ${\bf q}$ is the heat flux, given by 
\begin{equation}
{\bf q} = -\kappa_0T^{5/2}\nabla T
\end{equation}
with $\kappa_0 = 9\times 10^{-7}$ erg K$^{-7/2}$ cm$^{-1}$ s$^{-1}$. We can thus estimate the conductive cooling as
\begin{equation}
H_q = \frac{(\gamma -1)\kappa_0}{n k}\frac{T^{7/2}}{L^2}.
\end{equation}
To estimate $H_q$, we derive the density from the calculated emission measure using our characteristic path length of $l=10^9$ cm and take $L$ to be $10^{10}$ cm, which is the characteristic height of the supra-arcade fan.  The resulting estimated conductive cooling rate is plotted for three times in Figure \ref{cond_cooling.fig}.

Figure \ref{cond_cooling.fig} shows that at early times, around 12:07 UT, the conductive cooling rate in the 
central part of the supra-arcade region is fairly high, at about 0.1 MK/s.  This region of high conductive 
losses shifts to the lower right with time, and this shift corresponds to new SADs appearing in the lower right 
part of the supra-arcade region (i.e. SADs 11 and 12).  Referring to Figure \ref{heating_values.fig}, it seems that the adiabatic heating supplied by the
movement of the SADs is enough to overcome the losses due to the conductive cooling.

Figure \ref{cond_cooling.fig} also shows that the conductive cooling rate in the left part of the supra-arcade 
region, where there are no SADs, is on the order of 0.04 MK/s.  Referring to Figure \ref{temp_trend.fig}, Line 1 
in the no-SAD region cools the fastest, cooling from about 9.5 MK to 7.8 MK over the course of the 30 minutes 
from 12:00 UT to 12:30 UT.  These numbers give an empirical cooling rate of about 0.0009 MK/s.  This rate is 
much slower than the estimated conductive cooling rate.  We speculate that there may be some other global 
heating effect in this region that is serving to partially counteract the conductive cooling, such as a large 
scale compression of the plasma due to reconnection inflow, or ohmic dissipation in the current sheet.  It may also be the case that downflows do indeed exist in these lanes, but are not as easily detectable with the current AIA observations for a variety of reasons (e.g., speeds, relative intensities, sizes). These undetected flows may not be disturbing as much fan plasma, thereby not creating noticeable SADs at the spatial resolution of AIA. This situation would have the added consequence of less thermalization of the surrounding plasma through compression compared to that imparted by the more potent, detectable downflows.

\section{Conclusions \label{conclusions.sec}}

In this paper, we examine the thermal structure of the supra-arcade plasma sheet that forms during a flare 
and CME that occurred on 2011 October 22.  We find that in regions where there are not any supra-arcade 
downflows, the plasma either stays the same temperature or cools slightly over the half-hour from 12:00 UT to 
12:30 UT.  In the region where there are SADs, the plasma sunward of the SADs tends to be slightly heated.  We 
calculate the velocity field in the supra-arcade plasma and use it to determine the heating rates due to 
adiabatic compression and viscous heating.  We find that the adiabatic compression gives heating rates that are 
enough to overcome the conductive cooling that would naturally occur in plasma at a temperature of $\sim$10 MK.  
We also find that in the region where there are no detectable SADs, the plasma is cooling at a slower rate than one would 
expect given the conductive cooling.  This result indicates that there is possibly some other global heating mechanism 
affecting the plasma (such as large scale compressions or ohmic dissipation in the current sheet) or that smaller unobserved downflows exist below the instrumental resolution and are heating the plasma, but at a smaller rate than the detectable downflows.  Imaging instrumentation in hot bandpasses with spatial resolution better than AIA's would be extremely useful in disentangling these two potential scenarios.

 In the future, we would like to conduct a similar analysis on a larger sample of flares to see if the results of the analysis of this event are more broadly applicable.  We have identified three other flares observed with AIA that would be good candidates for this study, and we describe the flow fields of those flares in a separate paper \citep{Freed2016}.  We are also actively pursuing IRIS observations of the supra-arcade region in order to help with the accuracy of the flow field calculations.

\noindent\section*{Acknowledgements} 
 The authors thank the anonymous referee for comments that improved the paper. This work is supported by NASA grants NNX13AG54G, NNX14AD43G and NNX15AJ93G.  S. L. Savage is supported by the Hinode Project Office at Marshall Space Flight Center.  This work has benefited from the use of NASA's 
Astrophysics Data System.

\bibliographystyle{apj}

\end{document}